\documentclass{comnet}
\usepackage{graphicx}
\usepackage{xcolor}
\usepackage{listings}
\usepackage{amsmath} 
\usepackage{booktabs}
\usepackage{tabularx} 
\usepackage{algorithmic}
\usepackage{longtable}
\usepackage{subfig}
\usepackage{url}
\usepackage{array}
\usepackage{doi}
\usepackage{siunitx}
\usepackage[numbers]{natbib}
\hypersetup{
  colorlinks=false,
  pdfborder={0 0 0},
}
\begin{document}

\title{An Axiomatic Analysis of Path Selection Strategies for \\ Multipath Transport in Path-Aware Networks}

\shorttitle{Axiomatic Path Selection in Path-Aware Networks} 
\shortauthorlist{Alissa Baumeister, Sina Keshvadi} 

\author{
\name{Alissa Baumeister$^*$}
\address{University of Applied Sciences Luebeck, Lubeck, Germany \email{alissab@posteo.de}}
\and
\name{Sina Keshvadi}
\address{Thompson Rivers University, Kamloops, BC, Canada \email{$^*$Corresponding author: skeshvadi@tru.ca}}}
\maketitle

\begin{abstract}
{
    Path-aware networking architectures like SCION provide end-hosts with explicit control over inter-domain routing, while multipath transport protocols like MPTCP and MPQUIC enable the concurrent use of multiple paths. This combination promises significant gains in performance and policy enforcement, but it also creates a stark trade-off between individual performance optimization and overall network stability. This paper quantifies this trade-off through a rigorous axiomatic analysis. We evaluate a spectrum of algorithms, from greedy (Min-RTT) and cooperative (Round-Robin) to hybrid approaches (Epsilon-Greedy), against axioms of Efficiency, Loss Avoidance, Stability, and Fairness in a simulated path-aware environment.

    Our simulations reveal that purely greedy strategies, while efficient under low contention, induce catastrophic packet loss, increasing by over $>{}$18,000\% as the number of competing agents grow, due to herd effects that cause severe network instability. Conversely, cooperative strategies ensure fairness and stability but at the cost of underutilizing high-capacity paths. Crucially, we demonstrate that hybrid strategies resolve this conflict. The Epsilon-Greedy algorithm, for instance, achieves the highest efficiency of all tested strategies in high-contention scenarios while mitigating the instability inherent to the greedy approach. Our axiomatic analysis suggests that tunable, hybrid algorithms are essential for designing robust and high-performance path selection mechanisms for next-generation networks.
}
{Path-Aware Networks, Multipath Transport, Path Selection, Axiomatic Analysis, Network Simulation}
\end{abstract}

\section{Introduction}
\label{sec:introduction}

    The architecture of the public Internet, largely unchanged for decades, has traditionally treated the network as an opaque cloud. End-hosts have minimal control over the path their data traverses, relying on the collective decisions of routers running the Border Gateway Protocol (BGP). This paradigm is being challenged by the emergence of path-aware network architectures, with SCION (Scalability, Control, and Isolation On Next-Generation Networks)~\cite{perrig2017scion} standing out as a prominent example. By providing end-hosts with explicit, verifiable information about available inter-domain paths, SCION fundamentally shifts routing control from the network core to the edges, empowering applications to make intelligent, policy-driven pathing decisions.

    Concurrently, innovations at the transport layer provide the mechanisms to leverage this newfound path diversity. Protocols like Multipath QUIC (MPQUIC)~\cite{deconinck2017mpquic} and Multipath TCP (MPTCP)~\cite{ford2020mptcp} enable a single connection to operate over multiple network paths simultaneously. The synergy between a path-aware architecture like SCION and a multipath transport protocol like MPQUIC is profound: for the first time, end-hosts have both the granular visibility to select multiple paths and the mechanism to use them concurrently, promising significant gains in throughput, resilience, and connection quality.

    However, this powerful combination introduces a critical and poorly understood challenge: the path selection problem. When thousands or millions of independent agents are empowered to choose their own paths, their collective decisions can trigger complex systemic effects. A rational, locally optimal strategy for a single agent, such as greedily selecting the path with the lowest instantaneous latency, can lead to globally detrimental outcomes. This ``herd effect'' can cause cascading path overloads, wild traffic oscillations, severe packet loss, and profound unfairness, nullifying the theoretical benefits of multipath communication. The interaction between these dynamic end-host decisions and underlying congestion control algorithms creates a complex, coupled system whose emergent behavior is difficult to predict.

    In this paper, we address this challenge by introducing an axiomatic framework, inspired by axiomatizations of congestion control~\cite{zarchy2019axiomatizing} and end-host effects~\cite{scherrer2021axiomatic}, to evaluate algorithms against a holistic set of principles: \textbf{Efficiency}, which measures the utilization of network resources; \textbf{Loss Avoidance}, which quantifies the degree of congestion; \textbf{Stability}, which characterizes traffic volatility; and \textbf{Fairness}, which assesses the equitable distribution of bandwidth. We implement and evaluate a spectrum of strategies, ranging from simple greedy and cooperative baselines to more nuanced hybrid approaches that blend exploration with exploitation.

    Our findings provide clear, quantitative evidence of the fundamental trade-offs inherent in path selection and highlight the significant risks of simplistic strategies. This work makes the following key contributions:
    \begin{itemize}
        \item Quantification of the fundamental stability-efficiency trade-off in decentralized path selection, enabled by our principled axiomatic framework.
        \item Demonstration of catastrophic failure in greedy strategies due to herd effects, leading to a $>$18,000\% increase in packet loss under load.
        \item Identification of hybrid strategies as a superior compromise, with algorithms like Epsilon-Greedy achieving the highest network efficiency in high-contention scenarios while preventing the instability inherent to purely greedy approaches.
    \end{itemize}
    
    The remainder of this paper is organized as follows. Section~\ref{sec:related} discusses related work. Section~\ref{sec:methodology} details our simulation methodology, the algorithms under test, and the axiomatic framework. Section~\ref{sec:results} presents our quantitative results, followed by a discussion of their implications in Section~\ref{sec:discussion}. Finally, Section~\ref{sec:conclusion} concludes the paper and suggests directions for future work.
    
\section{Background and Related Work}
\label{sec:related}

This research is situated at the intersection of four key areas: multipath transport, path-aware architectures, path selection algorithms, and axiomatic network analysis.

\paragraph{Multipath Transport Protocols}
MPTCP~\cite{ford2020mptcp} and MPQUIC~\cite{deconinck2017mpquic} facilitate concurrent path usage, with packet schedulers optimizing data distribution. Schedulers like BLEST~\cite{ferlin2016blest} estimate blocking to minimize head-of-line issues in heterogeneous networks, while others like ECF~\cite{paasch2014ecf} prioritize earliest completion. These mechanisms focus on scheduling packets across a small set of given paths (e.g., Wi-Fi/LTE or IPv4/IPv6); our work addresses the upstream problem of how end-hosts should select that set of paths in a path-rich environment. While advanced multipath congestion controls like MPCC~\cite{gilad2020mpcc} use online learning and coupled congestion control~\cite{han2011coupled} ensures fairness, our work complements these efforts by examining the systemic impacts of the path selection logic itself.

\paragraph{Path-Aware Networking and SCION}
SCION~\cite{perrig2017scion} provides end-hosts with secure and scalable control over network paths, with prior work exploring its impact on availability~\cite{basin2011scion} and policy enforcement~\cite{king2020leveraging}. In such architectures, path selection is a decentralized, end-host-driven process. This differs fundamentally from the centralized traffic engineering common in Software-Defined Networks (SDN)~\cite{jain2013b4} or Content Delivery Networks (CDNs)~\cite{akamai}. Our investigation focuses on the collective dynamics that emerge from these decentralized decisions, an area largely unexplored in prior SCION research.

\paragraph{Path Selection Strategies}
Existing strategies span greedy (e.g., Min-RTT), cooperative (e.g., round-robin variants~\cite{handley2011path}), and hybrid approaches. Epsilon-greedy, for example, adapts multi-armed bandit algorithms~\cite{sutton2018rl} to balance exploration and exploitation. While the benefits of using multipath in conjunction with congestion control are known~\cite{handley2007path}, and recent work has modeled multipath routing~\cite{scaglione2013modeling} and selection in vehicular networks~\cite{chen2022vehicular}, there is no consistent framework for comparing their systemic effects. Our work aims to unify the evaluation of such strategies through a common set of axioms. The risk of herd effects in decentralized routing~\cite{chen2024rerouting} further underscores the need for this systemic analysis.

\paragraph{Axiomatic Analyses}
The use of axiomatic frameworks to formalize desired system properties is well-established, with notable applications in congestion control~\cite{zarchy2019axiomatizing} and in analyzing the performance effects of end-host path selection~\cite{scherrer2021axiomatic}. We adapt this approach to the domain of multipath strategies, defining a set of axioms relevant to network efficiency, stability, and fairness.

\section{Methodology}
\label{sec:methodology}

To systematically evaluate different path selection strategies, we developed a custom discrete-event simulator and an axiomatic framework for performance evaluation. The full source code for the simulator is available on GitHub~\cite{keshvadi2025} to facilitate reproducibility. This section details our simulation environment, the algorithms under test, and the metrics used to assess their impact.

\subsection{Simulation Environment}

Our custom simulator, developed in Python, models a set of competing agents transmitting data across a shared network topology. The environment is designed to be lightweight and flexible, allowing for the rapid evaluation of various algorithms and network configurations. Each simulation runs for 300 time steps, where each step represents 10\,ms to align with typical network latency scales and allow for fine-grained modeling of congestion dynamics.

\paragraph{Topology}
The network consists of a single source and destination connected by three parallel, independent paths. This configuration, defined in a JSON file, creates a simple yet non-trivial decision space with a clear trade-off between low-latency and high-capacity paths. This isolates the core dynamics of the selection algorithms from the confounding factors of complex routing. For the experiments in this paper, the paths were configured as follows:
\begin{itemize}
    \item \textbf{Path 1 (Low-Latency):} 50\,Mbps capacity, 20\,ms base RTT.
    \item \textbf{Path 2 (High-Capacity):} 100\,Mbps capacity, 50\,ms base RTT.
    \item \textbf{Path 3 (Balanced):} 80\,Mbps capacity, 80\,ms base RTT.
\end{itemize}

\paragraph{Agent and Transport Model}
The simulation is populated by $N$ agents, each representing an independent data flow with infinite demand (i.e., always having data to send). Each agent implements a simplified multipath transport mechanism and is governed by a standard Additive Increase, Multiplicative Decrease (AIMD) congestion controller, modeled after TCP Reno to provide a foundational behavioral baseline. This approach abstracts away the complexities of modern algorithms like CUBIC or BBR, allowing us to focus purely on the systemic effects induced by the path selection logic. The AIMD parameters are: initial congestion window (\textit{cwnd}) of 1 packet, additive increase $\alpha = 1$ packet per successful RTT, and multiplicative decrease $\beta = 0.5$ upon packet loss. At each time step, an agent sends data at a rate dictated by its \textit{cwnd}, with 1 unit of \textit{cwnd} normalized to 1\,Mbps.

\paragraph{Congestion Modeling}
Congestion is modeled at the path level. The simulator aggregates the total load from all agents on each path at each time step. If the aggregate load exceeds a path's capacity, a packet loss event is triggered for all agents using that path, with the loss amount proportional to their contribution to the overload. To model realistic RTT variations, the instantaneous RTT for a path is calculated as:
$$ \textit{RTT}_{\text{inst}} = \textit{RTT}_{\text{base}} + \max(0, k \times (\text{load} / \text{capacity} - 1)) $$
where $k = 10$\,ms is a scaling factor that simulates queuing delay.

\subsection{Path Selection Algorithms}

Each agent in a given simulation run employs the same path selection strategy. We evaluated the following seven algorithms:
\begin{enumerate}
    \item \textbf{Min-RTT (Greedy):} Selects the single path with the lowest measured instantaneous RTT for all traffic.
    \item \textbf{Min-Load (Cooperative):} Selects the path with the lowest current load. Our analysis assumes oracle-like knowledge of aggregate path loads, representing an idealized cooperative scenario.
    \item \textbf{Attribute-Aware:} A policy-driven strategy that first filters paths based on predefined attributes (e.g., avoiding a ``high-cost'' path) and then applies a secondary logic, which for these experiments was Min-RTT.
    \item \textbf{Round-Robin:} A stateless load-balancing algorithm that cycles through the available paths in a fixed sequence.
    \item \textbf{Weighted Round-Robin (WRR):} An enhanced load-balancing strategy where traffic is distributed proportionally to each path's configured capacity.
    \item \textbf{Epsilon-Greedy:} A hybrid strategy that balances exploitation and exploration. With probability $(1 - \epsilon)$, the agent exploits the best-known path (lowest RTT); with probability $\epsilon$, it explores by selecting a random path. For our experiments, $\epsilon=0.1$.
    \item \textbf{BLEST (Blocking Estimation):} A scheduler inspired by MPTCP literature~\cite{ferlin2016blest} that sends data over the path predicted to have the earliest completion time, factoring in latency and estimated queuing delay.
\end{enumerate}

\subsection{Axiomatic Framework for Performance Metrics}
\label{subsec:axioms}

To ensure a holistic comparison, we evaluate each simulation's outcome using a predefined set of performance axioms. These metrics are calculated from time-series data collected during each run.

\paragraph{Efficiency ($\eta$)} Measures the productive use of network resources, calculated as the average aggregate goodput (successful throughput after losses) across all agents and paths, in Mbps.

\paragraph{Loss ($\lambda$)} Quantifies the total volume of data dropped due to congestion. It is the sum of traffic load that exceeds path capacity at each time step, measured in Mbps.

\paragraph{Stability ($\sigma$)} Characterizes system volatility. We define stability as an inverse function of the mean standard deviation of the load across all paths over time (denoted \textit{oscillation}):
\begin{equation} \label{eq:stability}
\sigma = \frac{1}{1 + \textit{oscillation}}
\end{equation}
A high score (max 1.0) indicates a stable system, while a score near zero indicates high instability.

\paragraph{Fairness ($\phi$)} Evaluates the equity of bandwidth distribution among competing agents using Jain's Fairness Index on the final congestion window sizes of all $N$ agents. The index ranges from $1/N$ (worst case) to 1.0 (perfect fairness).

\paragraph{Loss Avoidance} A normalized score between 0 and 1 that provides an intuitive measure of a strategy's ability to prevent congestion, where 1 represents zero packet loss. It is calculated as:
\begin{equation} \label{eq:loss_avoidance}
\text{Loss Avoidance} = \frac{1}{1 + \lambda}
\end{equation}

\section{Results}
\label{sec:results}

To evaluate the performance of the path selection strategies, we conducted a series of discrete-event simulations. This section details the experimental setup, presents the quantitative results, and provides visual evidence of the key behavioral dynamics observed.

\subsection{Experimental Setup}

All experiments were conducted using our custom simulator, implemented in Python~\cite{keshvadi2025}. The primary network topology used for our analysis features three paths with a deliberate trade-off between latency and capacity: a low-latency, low-capacity path (Path 1: 20\,ms RTT, 50\,Mbps), a high-capacity, high-latency path (Path 2: 50\,ms RTT, 100\,Mbps), and a balanced path (Path 3: 80\,ms RTT, 80\,Mbps). Path 3 was also assigned a ``high-cost'' attribute for policy testing. We evaluated seven algorithms: Min-RTT, Min-Load, Attribute-Aware, Round-Robin, Weighted Round-Robin (WRR), Epsilon-Greedy (with $\epsilon=0.1$), and BLEST. For each algorithm, we ran simulations varying the number of competing agents from 10 to 500. Performance was evaluated using our axiomatic framework (Efficiency, Loss, Stability, Fairness, and Loss Avoidance), as defined in Section~\ref{subsec:axioms}.

\subsection{Performance Analysis}

The results reveal a clear and significant divergence in behavior between the different classes of algorithms. A high-level summary of the axiomatic scores for key scenarios is presented in Table~\ref{tab:summary_results}.

\begin{table}[htbp]
\centering
\caption{Axiomatic Performance Summary for Key Scenarios.}
\label{tab:summary_results}
\sisetup{round-mode=places, round-precision=2, table-format=4.2}
\begin{tabular}{l S[table-format=3.0] S S S[table-format=1.2] S[table-format=1.2]}
\toprule
\textbf{Strategy} & {\textbf{Agents}} & {\textbf{Efficiency} ($\eta$)} & {\textbf{Loss} ($\lambda$)} & {\textbf{Fairness} ($\phi$)} & {\textbf{Stability} ($\sigma$)} \\
\midrule
Min-RTT           & 10                & 44.79                        & 2.44                       & 0.33                        & 0.19                       \\
                  & 100               & 100.95                       & 50.37                      & 0.34                        & 0.15                       \\
                  & 500               & 504.84                       & 454.24                     & 0.34                        & 0.03                       \\
\midrule
Min-Load          & 10                & 60.01                        & 0.09                       & 0.33                        & 0.16                       \\
                  & 100               & 101.02                       & 20.52                      & 0.34                        & 0.14                       \\
                  & 500               & 504.58                       & 424.08                     & 0.34                        & 0.04                       \\
\midrule
Attribute-Aware   & 10                & 44.72                        & 2.42                       & 0.33                        & 0.19                       \\
                  & 100               & 100.92                       & 50.38                      & 0.34                        & 0.16                       \\
                  & 500               & 504.94                       & 454.34                     & 0.34                        & 0.03                       \\
\midrule
Round-Robin       & 10                & 20.05                        & 0.00                       & 1.00                        & 0.10                       \\
                  & 100               & 200.26                       & 123.13                     & 1.00                        & 0.01                       \\
                  & 500               & 1001.58                      & 924.42                     & 1.00                        & 0.00                       \\
\midrule
WRR               & 10                & 59.93                        & 0.92                       & 0.90                        & 0.03                       \\
                  & 100               & 206.47                       & 24.52                      & 0.92                        & 0.03                       \\
                  & 500               & 536.20                       & 313.57                     & 0.97                        & 0.01                       \\
\midrule
Epsilon-Greedy    & 10                & 41.46                        & 1.05                       & 0.36                        & 0.18                       \\
                  & 100               & 113.90                       & 49.73                      & 0.43                        & 0.15                       \\
                  & 500               & 570.12                       & 451.71                     & 0.43                        & 0.04                       \\
\midrule
BLEST             & 10                & 44.76                        & 2.43                       & 0.33                        & 0.19                       \\
                  & 100               & 100.90                       & 50.30                      & 0.34                        & 0.15                       \\
                  & 500               & 504.53                       & 453.93                     & 0.34                        & 0.04                       \\
\bottomrule
\end{tabular}
\end{table}

\subsubsection{The Herd Effect of Greedy Strategies}
Our results provide a cautionary tale: seemingly rational, greedy path selection is fundamentally unstable. As shown in Table~\ref{tab:summary_results}, this ``herd effect'' causes Min-RTT's packet loss to explode by over 18,000\% as the agent count increases from 10 to 500, transforming a seemingly high-performance system into one of catastrophic collapse. The loss for Min-RTT increases from 2.44\,Mbps with 10 agents to 454.24\,Mbps with 500 agents, causing its Loss Avoidance score to plummet to near-zero (0.002).

\begin{figure}[t!]
\centering
\includegraphics[width=\textwidth]{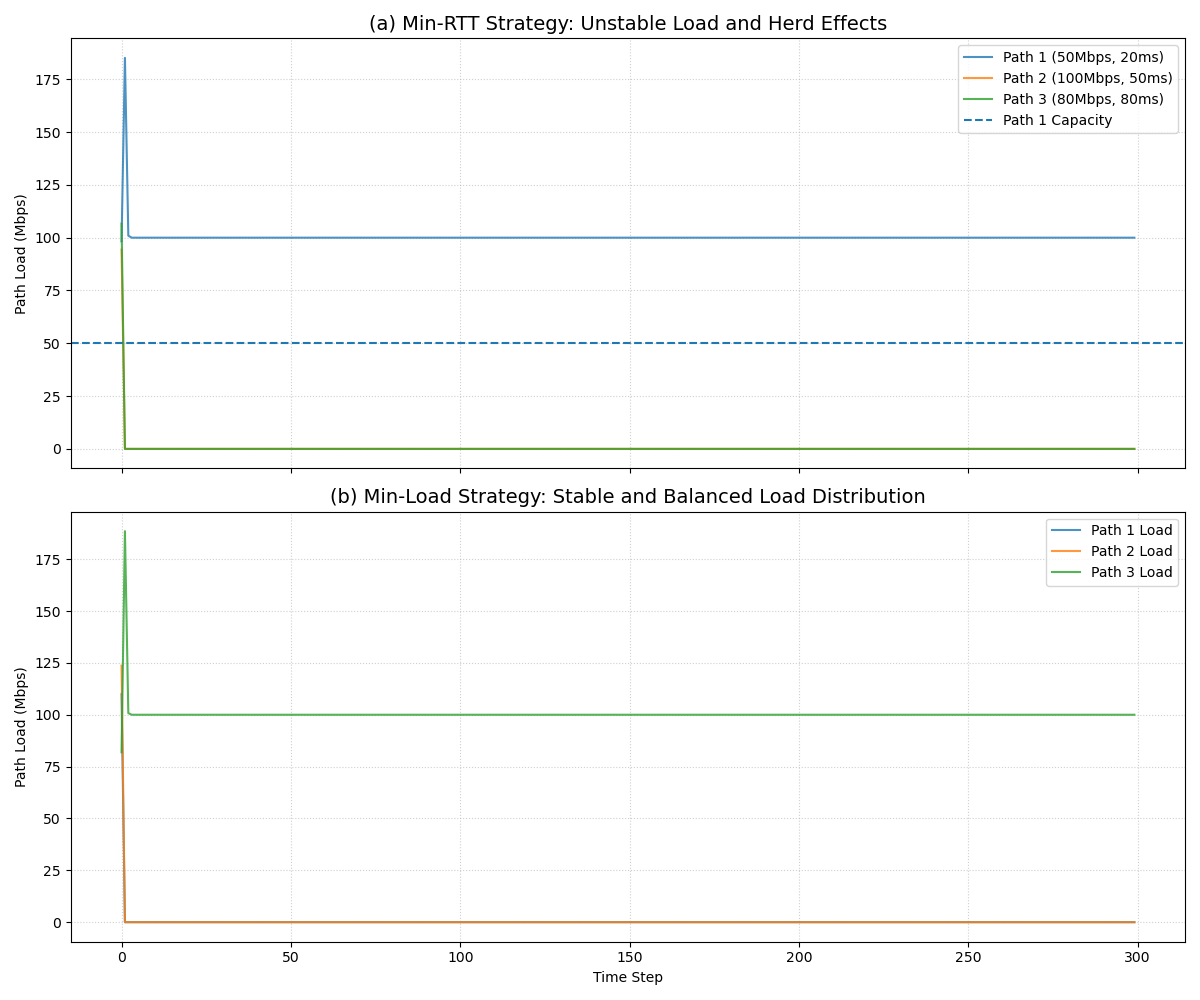}
\caption{Comparison of path load dynamics for (a) the greedy Min-RTT strategy and (b) the cooperative Round-Robin strategy with 100 agents. The sharp, synchronized oscillations in (a) illustrate the ``herd effect,'' where agents repeatedly overload the low-latency path (Path 1), causing self-inflicted congestion. In contrast, the stable, distributed loads in (b) demonstrate Round-Robin's ability to avoid congestion at the cost of suboptimal path utilization.}
\label{fig:path_load_dynamics}
\end{figure}

This phenomenon is visually represented in Figure~\ref{fig:path_load_dynamics}a. The load on the low-latency path is extremely volatile, characterized by sharp oscillations that frequently breach the path's capacity. This is a classic herd effect, where all agents independently select the same momentarily best path, causing a self-inflicted congestion collapse. Notably, the BLEST strategy exhibited nearly identical performance. In our topology, its initial filtering step---which prioritizes paths within 1.5x of the best RTT---consistently eliminated the higher-latency paths from consideration, causing its behavior to converge with the purely greedy algorithm.

\subsubsection{The Stability of Cooperative Strategies}
In stark contrast, cooperative strategies like Round-Robin and Min-Load prioritize stability and fairness. As seen in Table~\ref{tab:summary_results}, Round-Robin achieves a near-perfect Fairness score ($\phi \approx 1.00$) and incurs zero packet loss in low-contention scenarios. This stability is illustrated in Figure~\ref{fig:path_load_dynamics}b, where path loads are smoothly distributed, avoiding the chaotic spikes of the greedy approach. However, this stability comes at a significant performance cost. While Round-Robin's raw throughput appears high under contention, it is achieved with catastrophic loss (e.g., 924.42\,Mbps at 500 agents), making its effective goodput poor compared to more adaptive algorithms.

\subsubsection{Policy Enforcement with Attribute-Aware Selection}
The Attribute-Aware strategy successfully demonstrated its policy-enforcement capability. As designed, it completely avoided the path marked ``high-cost'' (Path 3). Since its secondary selection criterion was Min-RTT, its performance characteristics mirrored the purely greedy strategy. This confirms that network policy can be effectively layered on top of a performance-oriented selection logic.

\subsubsection{Hybrid Strategies as a Superior Compromise}
Our results show that hybrid strategies offer a compelling balance between the extremes of pure greed and pure cooperation. The Weighted Round-Robin (WRR) strategy, for example, biases traffic towards higher-capacity paths and significantly improves upon its simpler counterpart. At 500 agents, WRR achieves a network Efficiency of 536.20, outperforming Min-RTT (504.84), while inducing substantially less loss than the standard Round-Robin algorithm.

\begin{figure}[t!]
\centering
\includegraphics[width=\textwidth]{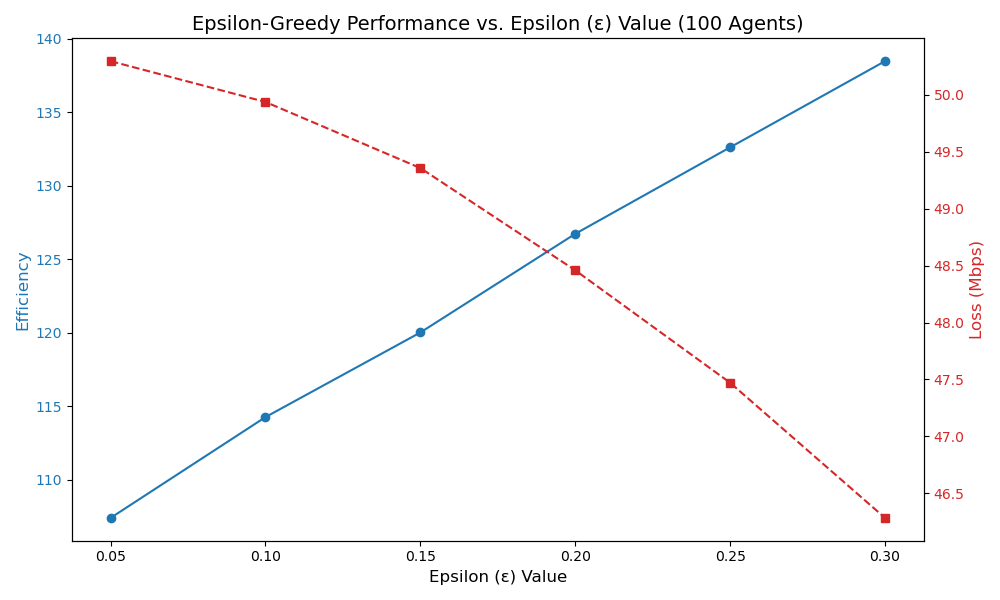}
\caption{Sensitivity analysis of the Epsilon-Greedy algorithm as a function of the exploration factor, $\epsilon$, with 500 agents. The plot demonstrates a clear, tunable trade-off: increasing $\epsilon$ (x-axis) reduces catastrophic packet loss ($\lambda$, red line) by breaking the herd effect, at the cost of a modest reduction in peak Efficiency ($\eta$, blue line).}
\label{fig:epsilon_sensitivity}
\end{figure}

The Epsilon-Greedy strategy proved particularly effective, achieving the highest Efficiency of all tested algorithms (570.12) at 500 agents. The small probability of random exploration ($\epsilon=0.1$) is sufficient to break the herd-effect symmetry, allowing the system to achieve higher performance without collapsing. To further analyze this, we conducted a sensitivity analysis on the $\epsilon$ parameter, with results shown in Figure~\ref{fig:epsilon_sensitivity}. The plot reveals a clear, tunable trade-off: as $\epsilon$ increases, packet Loss decreases dramatically. The cost of this increased stability is a modest and graceful reduction in peak Efficiency. This demonstrates that hybrid algorithms can be tuned to meet specific performance goals, balancing the need for high throughput with the imperative of network stability.

\section{Discussion}
\label{sec:discussion}

Our results reveal fundamental trade-offs in decentralized path selection and provide clear insights for designing effective multipath transport strategies. In this section, we interpret these findings, discuss their broader implications, and acknowledge the limitations of our study.

\subsubsection{Interpretation of Key Findings}
A core finding of this work is that in a competitive, decentralized environment, no simple path selection strategy is optimal across all performance dimensions. We identified two primary classes of behavior---efficiency-seeking (greedy) and stability-seeking (cooperative)---each with significant drawbacks. The catastrophic failure of the Min-RTT strategy under load is not a flaw in the algorithm itself but an emergent property of the system: a ``tragedy of the commons'' where rational, individual decisions lead to a collectively poor outcome. The herd effect, visualized in Figure~\ref{fig:path_load_dynamics}a, demonstrates that without coordination, agents perpetually rush to, overload, and subsequently flee the best available resource in a synchronized cycle, leading to high packet loss and severe instability.

\subsubsection{The Case for Hybrid Strategies}
The superior performance of the Epsilon-Greedy and Weighted Round-Robin strategies stems from their ability to implicitly coordinate agent behavior. The random ``exploration'' step in Epsilon-Greedy acts as a crucial symmetry-breaking mechanism. It forces a small fraction of agents to deviate from the herd, effectively discovering and utilizing otherwise empty paths. This minimal exploration is sufficient to prevent the catastrophic overload cycles of a purely greedy approach. As shown in our sensitivity analysis (Figure~\ref{fig:epsilon_sensitivity}), this behavior is tunable, allowing for a direct trade-off between maximizing efficiency and minimizing loss. Similarly, WRR leverages static path information (capacity) to inject a form of \textit{a priori} knowledge into the system, leading to a more globally efficient outcome than a naive Round-Robin scheduler.

\subsubsection{Implications for Network Designers and Operators}
Our results have several practical implications for the design of future multipath protocols and path-aware networks:
\begin{itemize}
    \item \textbf{Avoid Purely Latency-Based Selection:} Schedulers for protocols like MPQUIC should not rely solely on latency as a selection metric, as this is demonstrably unstable under load.
    \item \textbf{Embrace Tunable Randomness:} Incorporating a small, tunable degree of randomness or exploration into path selection is a powerful and lightweight tool for ensuring system stability without a significant performance penalty.
    \item \textbf{Leverage Path Attributes:} The success of the Attribute-Aware and WRR strategies suggests that network providers in a SCION-like architecture should expose rich path metadata. Attributes indicating capacity, utilization, or cost can help end-hosts make more intelligent and globally beneficial decisions.
\end{itemize}

\subsection{Limitations and Future Work}
Our study has several limitations that present clear avenues for future work. First, our simulator employs a simplified AIMD congestion controller; the system dynamics may differ with more modern algorithms like CUBIC or BBR. Second, our analysis primarily used a single, albeit representative, network topology. While we validated our findings on a second topology, real-world network structures are far more diverse. Finally, our simulation represents a closed world with homogeneous agents and no external cross-traffic, which could affect algorithm performance in a real-world deployment.

\subsection{Data and Code Availability}
The source code for the discrete-event simulator, the JSON configuration files for all topologies, and the complete raw and summarized result datasets are available in a public GitHub repository~\cite{keshvadi2025}.

\section{Conclusion}
\label{sec:conclusion}

The transition to path-aware networking architectures presents a profound opportunity to enhance the performance and reliability of Internet communication. However, empowering end-hosts with path selection capabilities introduces significant challenges related to systemic stability. In this paper, we conducted a rigorous axiomatic analysis of decentralized path selection strategies, establishing a clear framework for evaluating their performance and trade-offs.

Our findings quantitatively demonstrate a fundamental trade-off between efficiency and stability. We show that simple, greedy strategies like Min-RTT are prone to catastrophic failure under load due to herd effects, while purely cooperative strategies are stable but inefficient. The central contribution of this work is the demonstration that hybrid strategies provide a superior compromise. Algorithms like Epsilon-Greedy, which blend greedy exploitation with a small degree of random exploration, can successfully break the symmetry of herd behavior, achieving high network efficiency while mitigating the worst effects of congestion and instability.

This work underscores the critical need for intelligent, hybrid schedulers in future multipath transport protocols. Future research should extend this analysis by incorporating more sophisticated congestion control algorithms and exploring a wider range of network topologies. Ultimately, validating these simulation findings on a real-world SCION testbed will be an essential next step.

\section*{Acknowledgments}

This work was supported by Globalink Mitacs and DAAD. We also acknowledge the use of generative AI tools for assistance with grammar, phrasing, and code refinement.

\bibliographystyle{comnet}
\bibliography{main}
\appendix
\section{Full Experimental Results}
\label{sec:appendix_full_results}
This section contains the unabridged summary table of axiomatic performance metrics for all tested strategies across all agent counts.
\begingroup 
\small 
\setlength{\tabcolsep}{4pt} 

\begin{longtable}{llrrrrrr}
\caption{Axiomatic Performance Summary for All Strategies}
\label{tab:full_results} \\
\toprule
Strategy & Agents & Oscillation &    Loss & Fairness & Efficiency & Stability & Loss Avoidance \\
\midrule
\endfirsthead
\caption[]{Axiomatic Performance Summary for All Strategies (Continued)} \\
\toprule
Strategy & Agents & Oscillation &    Loss & Fairness & Efficiency & Stability & Loss Avoidance \\
\midrule
\endhead
\midrule
\multicolumn{8}{r}{{Continued on next page}} \\
\midrule
\endfoot
\bottomrule
\endlastfoot
min\_rtt & 10 & 4.25 & 2.44 & 0.33 & 44.79 & 0.19 & 0.29 \\
min\_rtt & 25 & 5.58 & 0.43 & 0.34 & 38.42 & 0.15 & 0.70 \\
min\_rtt & 50 & 10.04 & 24.81 & 0.34 & 75.11 & 0.09 & 0.04 \\
min\_rtt & 100 & 5.51 & 50.37 & 0.34 & 100.95 & 0.15 & 0.02 \\
min\_rtt & 150 & 8.38 & 100.78 & 0.34 & 151.38 & 0.11 & 0.01 \\
min\_rtt & 250 & 12.90 & 201.71 & 0.34 & 252.31 & 0.07 & 0.00 \\
min\_rtt & 500 & 28.00 & 454.24 & 0.34 & 504.84 & 0.03 & 0.00 \\
min\_load & 10 & 5.25 & 0.09 & 0.33 & 60.01 & 0.16 & 0.92 \\
min\_load & 25 & 7.86 & 6.59 & 0.33 & 74.89 & 0.11 & 0.13 \\
min\_load & 50 & 10.64 & 10.21 & 0.34 & 75.63 & 0.09 & 0.09 \\
min\_load & 100 & 6.23 & 20.52 & 0.34 & 101.02 & 0.14 & 0.05 \\
min\_load & 150 & 8.91 & 70.95 & 0.34 & 151.45 & 0.10 & 0.01 \\
min\_load & 250 & 14.67 & 151.94 & 0.34 & 252.37 & 0.06 & 0.01 \\
min\_load & 500 & 26.84 & 424.08 & 0.34 & 504.58 & 0.04 & 0.00 \\
attribute\_aware & 10 & 4.30 & 2.42 & 0.33 & 44.72 & 0.19 & 0.29 \\
attribute\_aware & 25 & 5.43 & 0.44 & 0.34 & 38.42 & 0.16 & 0.70 \\
attribute\_aware & 50 & 10.40 & 24.99 & 0.34 & 75.31 & 0.09 & 0.04 \\
attribute\_aware & 100 & 5.09 & 50.38 & 0.34 & 100.92 & 0.16 & 0.02 \\
attribute\_aware & 150 & 8.74 & 100.90 & 0.34 & 151.50 & 0.10 & 0.01 \\
attribute\_aware & 250 & 14.04 & 201.75 & 0.34 & 252.35 & 0.07 & 0.00 \\
attribute\_aware & 500 & 28.33 & 454.34 & 0.34 & 504.94 & 0.03 & 0.00 \\
round\_robin & 10 & 9.44 & 0.00 & 1.00 & 20.05 & 0.10 & 1.00 \\
round\_robin & 25 & 23.63 & 0.08 & 1.00 & 50.16 & 0.04 & 0.93 \\
round\_robin & 50 & 47.21 & 23.36 & 1.00 & 100.28 & 0.02 & 0.04 \\
round\_robin & 100 & 94.27 & 123.13 & 1.00 & 200.26 & 0.01 & 0.01 \\
round\_robin & 150 & 141.44 & 223.29 & 1.00 & 300.46 & 0.01 & 0.00 \\
round\_robin & 250 & 235.70 & 423.52 & 1.00 & 500.69 & 0.00 & 0.00 \\
round\_robin & 500 & 471.43 & 924.42 & 1.00 & 1001.58 & 0.00 & 0.00 \\
weighted\_round\_robin & 10 & 29.21 & 0.92 & 0.90 & 59.93 & 0.03 & 0.52 \\
weighted\_round\_robin & 25 & 31.16 & 2.91 & 0.85 & 115.24 & 0.03 & 0.26 \\
weighted\_round\_robin & 50 & 23.57 & 6.21 & 0.88 & 168.98 & 0.04 & 0.14 \\
weighted\_round\_robin & 100 & 28.69 & 24.52 & 0.92 & 206.47 & 0.03 & 0.04 \\
weighted\_round\_robin & 150 & 31.36 & 40.90 & 0.93 & 235.25 & 0.03 & 0.02 \\
weighted\_round\_robin & 250 & 38.93 & 93.34 & 0.96 & 303.45 & 0.03 & 0.01 \\
weighted\_round\_robin & 500 & 109.03 & 313.57 & 0.97 & 536.20 & 0.01 & 0.00 \\
epsilon\_greedy & 10 & 4.41 & 1.05 & 0.36 & 41.46 & 0.18 & 0.49 \\
epsilon\_greedy & 25 & 6.70 & 4.16 & 0.39 & 48.63 & 0.13 & 0.19 \\
epsilon\_greedy & 50 & 9.75 & 16.68 & 0.41 & 73.15 & 0.09 & 0.06 \\
epsilon\_greedy & 100 & 5.71 & 49.73 & 0.43 & 113.90 & 0.15 & 0.02 \\
epsilon\_greedy & 150 & 9.05 & 100.10 & 0.43 & 171.27 & 0.10 & 0.01 \\
epsilon\_greedy & 250 & 13.27 & 200.56 & 0.43 & 284.38 & 0.07 & 0.00 \\
epsilon\_greedy & 500 & 25.11 & 451.71 & 0.43 & 570.12 & 0.04 & 0.00 \\
blest & 10 & 4.16 & 2.43 & 0.33 & 44.76 & 0.19 & 0.29 \\
blest & 25 & 5.14 & 0.49 & 0.34 & 38.48 & 0.16 & 0.67 \\
blest & 50 & 10.08 & 24.83 & 0.34 & 75.13 & 0.09 & 0.04 \\
blest & 100 & 5.52 & 50.30 & 0.34 & 100.90 & 0.15 & 0.02 \\
blest & 150 & 8.20 & 100.77 & 0.34 & 151.37 & 0.11 & 0.01 \\
blest & 250 & 13.89 & 201.68 & 0.34 & 252.28 & 0.07 & 0.00 \\
blest & 500 & 26.26 & 453.93 & 0.34 & 504.53 & 0.04 & 0.00 \\
\end{longtable}

\endgroup 

\end{document}